# Intrinsic Correlated Electronic Structure of $CrO_2$ Revealed by Hard X-ray Photoemission Spectroscopy

M. Sperlich[1*], C. König[1], G. Güntherodt[1+], A. Sekiyama[2], G. Funabashi[2], M. Tsunekawa[2], S. Imada[2], A. Shigemoto[2], K. Okada[3], A. Higashiya[4], M. Yabashi[4], K. Tamasaku[4], T. Ishikawa[4], V. Renken[5], T. Allmers[5], M. Donath[5] and S. Suga[2]

[1] *II. Physikalisches Institut, RWTH Aachen University, 52074 Aachen, Germany*

[2] *Graduate School of Engineering Science, Osaka University, Toyonaka, Osaka 560-8531, Japan*

[3] *Graduate School of Natural Science and Technology, Okayama University, Okayama 700-8530, Japan*

[4] *SPring-8/Riken, 1-1-1 Kouto, Mikazuki, Sayo, Hyogo 679-8148, Japan*

[5] *Physikalisches Institut, Westfälische Wilhelms-Universität Münster, 48149 Münster, Germany*

Abstract

Bulk-sensitive hard x-ray photoemission spectroscopy (HAXPES) reveals for as-grown epitaxial films of half-metallic ferromagnetic $CrO_2$(100) a pronounced screening feature in the Cr $2p_{3/2}$ core level and an asymmetry in the O 1*s* core level. This gives evidence of a finite, metal-type Fermi edge, which is surprisingly not observed in HAXPES. A spectral weight shift in HAXPES away from the Fermi energy is attributed to single-ion recoil effects due to high energy photoelectrons. In conjunction with inverse PES the intrinsic correlated Mott-Hubbard-type electronic structure is unravelled, yielding an averaged Coulomb correlation energy $U_{av} \cong 3.2$ eV.

Transition metal oxides are strongly correlated electron systems, which exhibit a wealth and future potential of phenomena most challenging to modern solid state physics [1, 2]. In the theoretical description of the electronic structure of transition metal oxides a seminal progress is owed to dynamic mean-field theory (DMFT) [3,4]. This method has called upon intrinsic, bulk-sensitive photoemission spectroscopy (PES) [5]. The reason is obvious because electronic states of the clean surface differ from those in the bulk due to the increase in $U/t$, where $U$ is the on-site electron Coulomb repulsion energy and $t$ is the electron hopping energy between lattice sites. Experimentally a breakthrough toward determining the intrinsic bulk electronic structure occurred due to the development of hard x-ray PES (HAXPES) with a probing depth of 5 – 10 nm [6 – 9]. In this context a very overdue, controversial and provoking case is the half-metallic ferromagnet $CrO_2$ [10,11], which exhibits a metastable surface, transforming into the stable antiferromagnetic insulator $Cr_2O_3$ [10]. For $CrO_2$ a discrepancy exists between the correlated Fermi liquid-type metallic behavior [10,13] and the very small intensity of the sputter-cleaned surface in ultraviolet PES (UPES) near the Fermi energy $E_F$ [12]. The latter was conjectured to be due to surface relaxation of $CrO_2(001)$ [14]. The metallicity of $CrO_2(100)$ was even questioned based on UPES measurements [15]. On the contrary, an enhanced spectral weight near $E_F$ due to an orbital Kondo effect has been predicted using DMFT [16]. However, the theoretical description of electronic and (magneto-)optical data of $CrO_2$ has raised doubts about the relevance of strong Hubbard-type correlations [17-20]. This controversy and the above surface calamity stress the indispensable need to employ HAXPES in comparison to soft x-ray PES (SXPES) [8,21,22]. Despite the interest in $CrO_2$ for spintronics applications [23,24], because of its high spin polarization [12,17,25-28], the intrinsic correlated electronic structure still remains to be unravelled [29].

Here we present the first bulk-sensitive investigation of valence band states and core levels of $CrO_2$ by means of HAXPES using photon energies of $h\nu \approx 8$ keV. The photoemission intensity near $E_F$ observed by HAXPES is unexpectedly small, in contrast to the metal-type Fermi edge observed by SXPES. However, by HAXPES we found a strong metallic screening feature in the Cr $2p_{3/2}$ core level and an asymmetry of the O $1s$ core level, which both imply a finite density of states (DOS) near $E_F$. This seeming contradiction with the very small photoemission intensity near $E_F$ in HAXPES is resolved by considering single-ion recoil effects in HAXPES. They account for the suppression of the spectral weight near $E_F$ due to its shift to higher binding energy (BE). Using HAXPES, SXPES and inverse PES (IPES), we identify the salient intrinsic features of the correlated Mott-Hubbard-type electronic structure of $CrO_2$.

A correlated electronic structure of $CrO_2$ has been concluded from calculations using local spin density approximation LSDA + $U$ [17]. The Cr $3d$ states split in the octahedral crystal field into a lower and upper state with $t_{2g}$ and $e_g$ symmetry, respectively. The $t_{2g}$ states split further into a strongly localized $3d$(xy) orbital near 1 eV BE below $E_F$ and more dispersive $3d$(yz±zx) orbitals. The latter are strongly hybridized with the O $2p$ states, forming bands which cross $E_F$ and cause a self-doping of $CrO_2$ [17]. The exchange splitting shifts the minority spin states above $E_F$, giving rise to a spin gap.

The HAXPES and SXPES experiments have been performed at 150 K or 20 K at the BL19LXU and BL25SU beam lines of SPring-8 [30], respectively; we used as-grown, otherwise untreated surfaces of $CrO_2$(100) epitaxial films. The samples were grown by chemical vapor deposition in an oxygen atmosphere on (100)-oriented $TiO_2$ substrates [23,28,30]. To overcome the low photoionization cross section of the Cr $3d$ and O $2p$ valence states for $h\nu > 1000$ eV the PES resolution was set to 250 meV (FWHM), while it was set to 100 meV, 60 meV and 20 meV, respectively, for $h\nu = 700$ eV, 200 eV and 11.6 eV. To optimize the photoelectron emission, a so-called p-polarization configuration was employed for HAXPES, whereas fully circularly polarized light was used for SXPES below $h\nu = 2$ keV. The UPES ($h\nu = 11.6$ eV) and IPES measurements have been performed in laboratory systems, where both the as-grown and sputter-cleaned surface was measured [30].

Figure 1 shows valence band PES of as-grown $CrO_2(100)$ in normal emission at different photon energies (and varying probing depth $\delta$ [31]), ranging from 11.6 eV ($\delta \leq 1$ nm) to 8180 eV ($\delta \sim 10$ nm). Besides a broad peak near 1.75 eV BE for $h\nu = 200$ eV, a peak emerges with increasing photon energy near 1.0 eV BE which becomes enhanced for $h\nu = 8180$ eV. At lower photon energies (< 200 eV) the contribution of the insulating surface $Cr_2O_3$-layer of roughly 2 nm thickness [24] is mostly probed judging from the low photoemission intensity near $E_F$ (Fig. 1 and Ref. 32). With higher photon energies of 700 eV ($\delta \sim 1.4$ nm) and 1220 eV ($\delta \sim 2.2$ nm) the photoemission spectra show a metal-type Fermi edge. The Fermi edge for $h\nu = 1220$ eV is broader than the one for $h\nu = 700$ eV, because of its lower resolution. A metallic Fermi edge was also observed for $h\nu = 385$ eV with 400 meV resolution [33]. However, most unexpectedly we observe no metal-like Fermi edge for HAXPES using $h\nu = 8180$ eV, for which bulk properties are expected. Please note that for $E_F \leq BE \leq 0.2$ eV the very weak intensity increases almost linearly with increasing BE, showing a steeper slope above 0.2 eV BE.

In Fig. 2 we show the valence band spectrum of $CrO_2$ for $h\nu = 7942$ eV and for $E_F \leq BE \leq 3$ eV at 20 K and 150 K together with the Fermi edge of Au. Besides the prominent peak at 1.0 eV BE a shoulder near 2.1 eV BE is identified. Please note that there is no significant temperature dependence in the spectra in Fig. 2. The photoemission intensity for 20 K near $E_F$ is expanded by a factor of 5, differing strongly from that of Au. It evidences the absence of a metallic Fermi edge of $CrO_2$ in HAXPES. The features of Fig. 2 and its overview for $E_F \leq BE \leq 14$ eV [30] will be discussed below.

The HAXPES core level spectra help to resolve the puzzle about the intrinsic metallicity of $CrO_2$. The O $1s$ core level in Fig. 3(a) for $h\nu = 1490$ eV exhibits satellites at 3 eV and 6 eV BE above its maximum at 528 eV, which are strongly reduced for $h\nu = 7942$ eV and $h\nu = 8180$ eV. These satellites are due to O $2p$ – O $2p$ charge transfer and result most likely from a surface induced chemical shift of the O $1s$ core level of $Cr_2O_3$ and from surface adsorbed oxygen [30]. Most pronounced is the asymmetry in the line shape of the O $1s$ core level at 528 eV BE in the latter two HAXPES spectra. This asymmetry reflects the intrinsic finite DOS of unoccupied states near $E_F$ [34,35], implying also a non-zero O 2p and Cr 3d partial DOS of occupied states at $E_F$. The weak peaks in the HAXPES spectra near 9 eV and 11 eV BE above the O $1s$ peak are due to Cr $3d(e_g)$ – O $2p$ charge transfer satellites analogous to the case of cuprates [35]. The broad feature in Fig. 3(a) near 29 eV BE above the O $1s$ peak is attributed to plasmon excitations.

Another conclusive feature is found for the Cr $2p_{3/2}$ core level at 576 eV BE in Fig. 3(b). For $h\nu = 1490$ eV the Cr $2p_{3/2}$ level exhibits a weak shoulder near 575 eV BE, which develops for $h\nu \approx 8$ keV into a small but sharp peak. This sharp peak at 575 eV BE is obviously a bulk feature and is attributed to a well-screened satellite [8,34]. The metallic screening of the 2p core-hole potential in the PES final state is due to charge transfer from valence band states at $E_F$. Such a well-screened satellite was identified by HAXPES for the Mn $2p_{3/2}$ level in the metallic regime of $La_{1-x}Sr_xMnO_3$ [8]. This screening due to hybridized Mn-3d and doping-induced states near $E_F$ of metallic $La_{1-x}Sr_xMnO_3$ has to be replaced in the case of $CrO_2$ by the $2p$-$3d$ hybridized states near $E_F$. The latter states account for the well-screened feature of the $2p_{3/2}$ level of $CrO_2$, supporting its intrinsic metallicity for one spin channel. The Cr $2p_{1/2}$ level positioned at 586 eV BE does not exhibit such a satellite, most likely due to multiple configurational interactions and lifetime effects [30].

In order to reconcile the discrepancy in HAXPES between the well-screened $2p_{3/2}$ core level evidencing metallicity and the very small photoemission intensity near $E_F$, we attribute the latter observation to recoil effects [36], i.e. to a shift of orbital-dependent spectral weight to BE higher than $E_F$. These recoil effects induced by the emission of high energy photoelectrons in HAXPES are relevant not only upon photoexciting core levels, but also valence band states of light elements [37]. Instead of the whole crystal, the single ion is found to accept the photoelectron momentum. The single-ion recoil shift is given by $E_R \sim E_K \cdot (m/M) \sim (h\nu - BE) \cdot (m/M)$, where $E_K$ is the photoelectron kinetic energy, $m$ the electron mass and $M$ the nucleus mass. Estimates of the single-ion recoil shifts show that, e.g., the O 1$s$ core level of $CrO_2$ near 528 eV BE is shifted towards larger BE with respect to $E_F$ by $E_R^H \sim 260$ meV in HAXPES ($h\nu = 8180$ eV) compared to $E_R^S \sim 24$ meV in SXPES ($h\nu = 1220$ eV). The difference between the two types of spectra amounts to $E_R^{H-S}$(O 1$s$) = 236 meV. For the Cr $2p_{3/2}$ core level near 576 eV BE the recoil shift is estimated as $E_R^H \sim 81$ meV compared to $E_R^S \sim 7$ meV, thus yielding $E_R^{H-S}$(Cr $2p_{3/2}$) = 74 meV. Consequently, the splitting between O 1$s$ and Cr $2p_{3/2}$ states is estimated as $E_R^{H-S}$(O 1$s$) – $E_R^{H-S}$(Cr $2p_{3/2}$) = 236 meV - 74 meV = 162 meV smaller in HAXPES compared to SXPES. In our experiment, the O 1$s$ and Cr $2p_{3/2}$ core levels measured each for $h\nu = 7942$ eV with reference to $h\nu = 1490$ eV (Supplemental Fig. 4 [30]) show (H-S) shifts of 180 meV and 70 meV, respectively, which compare reasonably well with the corresponding estimates of 219 meV and 68 meV. Hence, the experimental splitting of $E_R^{H-S}$(O 1$s$) – $E_R^{H-S}$(Cr $2p_{3/2}$) = 180 – 70 = 110 meV is in fair agreement with the estimate of 219 – 68 = 151 meV, given the crude single-ion recoil approximation. On the other hand, concerning the 2$p$-3$d$ hybridized valence band states of $CrO_2$, which show experimentally a wide spread in energy and no specific lineshape, the recoil shifts cannot be determined straightforwardly. Hence an upper estimate of the recoil shifts between HAXPES ($h\nu = 8180$ eV) and SXPES ($h\nu = 1220$ eV) of the O 2$p$ and Cr 3$d$ states is obtained by comparison with the above O 1$s$ and Cr 2$p$ states, amounting at most to $E_R^{H-S}$(O 2$p$) - $E_R^{H-S}$(Cr 3$d$) $\approx$ 162 meV. Because of the experimental resolution, only a $p$- vs $d$-states weighted average of the two recoil shifts can be observed, which is estimated by these numbers to at least 135 meV [30]. This shift value agrees roughly with the energy range $E_F \leq$ BE $\leq 0.2$ eV in the HAXPES spectrum ($h\nu = 8180$ eV, Fig. 1) over which the intensity is increasingly suppressed toward smaller BE. This gradual suppression toward $E_F$ instead of a rigid shift is attributed to nonideal single nucleus recoil due to slight collisions with neighboring atoms. Please note that despite the 250 meV HAXPES resolution, a recoil shift of, e.g., 100 meV can still be resolved [30]. Moreover, the strongly suppressed intensity of the $p$-$d$ hybridized states for $E_F \leq$ BE $\leq 0.2$ eV in HAXPES compared to SXPES (Fig. 1) is not due to a more strongly decreasing photoionization cross section of the O 2$p$ atomic subshell with increasing $h\nu$ compared to the Cr 3$d$ subshell [30].

We now discuss the electronic structure of $CrO_2$ in terms of the Mott-Hubbard model [3,4,29,34]. Based on the above discussion we attribute the peak in Fig. 2 near 1 eV BE below $E_F$ to the (coherent) quasiparticle peak of $p$-$d$ hybridized states and the weak (incoherent) peak near 2.1 eV to the lower Hubbard band (LHB). The latter feature appeared similarly in UPES after prolonged surface sputtering [28]. The weak intensity near 2.1 eV in Fig. 2 is most likely due to the strong $p - d$ hybridization. A similarly weak photoemission intensity has been found for the LHB of strongly correlated $Sr_2RuO_4$ [38]. On the other hand, the upper Hubbard band (UHB) is identified by IPES [39]. In this experiment, the emitted photon energy is 9.9 eV and the overall energy resolution is 350 meV [40]. After sputtering the as-grown film surface, the structureless background intensity in Fig. 4 changes into a broad maximum around about -2.8 eV BE above $E_F$, which we attribute to the UHB. A small but finite intensity appears near $E_F$, consistent with UPES (h$\nu$ = 11.6 eV) after identical sputtering (Supplem. Fig. 1 [30]). A peak near -3.6 eV BE above $E_F$ has been found in "bremsstrahlung isochromat spectroscopy" [41]. The measurements, however, were carried out on compressed $CrO_2$ powder samples, which were scraped in situ. By the energy difference between the LHB near 2.1 eV and the UHB at |- 2.8 eV| we obtain an estimate of the local (intraorbital)

Coulomb repulsion $U$ of about $4.9 \pm 0.2$ eV. From this value we have to subtract the $d$-$d$ exchange splitting of 1.7 eV obtained from the LSDA calculation [25], which neglects correlation effects. Hence, we obtain for the averaged $d$-$d$ Coulomb interaction [5] $U_{av} \cong 4.9 - 1.7 \cong 3.2$ eV. A value of the $d$-$d$ Coulomb interaction $U = 3.0$ eV has been used in electronic structure calculations due to the constrain screening method [17].

In analogy to the screening of the Cr $2p_{3/2}$ core level it may be suggestive to consider the Kondo screening of a localized $d$ moment by the Cr $3d$ – O $2p$ hybridized states. Craco et al. [16] tested the scenario of an orbital Kondo effect within LDA + $U$ and DMFT. As impurity solver the IPT (iterated perturbation theory) approximation was used. A pronounced quasicoherent spectral weight is predicted at $E_F$. It disagrees, however, with the experimentally observed photoemission intensity near $E_F$ for photon energies ranging from 11.6 eV to 8180 eV and for temperatures between 300 K and 20 K (Figs. 1, 2 and Supplemental Figs. 1, 2 [30]).

In conclusion, the comparison of core level and valence state shifts of $CrO_2$ in HAXPES and SXPES reveals the crucial role of single-ion recoil effects in HAXPES. They result for the $p - d$ hybridized valence band states in a significant shift (> 100meV) of spectral weight toward higher BE below $E_F$. $CrO_2$ appears as favorable recoil-effect case because of its less dense, open rutile structure, where about 66 % of the unit cell volume lies outside the atomic spheres [25]. Despite the small HAXPES intensity near $E_F$, the asymmetry of the O $1s$ core level and the bulk-type screening feature in the Cr $2p_{3/2}$ level corroborate the metallicity of $CrO_2$ and its intrinsic non-vanishing, finite DOS near $E_F$. This is also supported by our IPES (Fig. 4) and UPES [30] measurements for the sputter-cleaned samples. Along this line we do not find evidence for an anomaly near $E_F$ due to a predicted orbital Kondo effect. Employing HAXPES and IPES, the intrinsic correlated Mott-Hubbard-type electronic structure is identified unambiguously. For the latter there remains still room for a deeper understanding by applying the DMFT approach [29].

**Acknowledgements**

We thank T. Saita, T. Kobayashi, T. Hamasaki, Y. Amano and Y. Yakubo for help with the SXPES and HAXPES measurements. S.S. acknowledges the Helmholtz Association and the A. von Humboldt Foundation for supporting his research stay in Germany from 2010 on. Fruitful discussions with D. Vollhardt are gratefully acknowledged.

* Present address**:** AREVA NP GmbH, Paul-Gossen-Straße 100, 91052 Erlangen, Germany
+ Corresponding author: gernot.guentherodt@physik.rwth-aachen.de

**References**

[1] Y. Tokura and N. Nagaosa, Science **288**, 462 (2000).

[2] Y. Tokura, Physics Today **56**(7), 50 (2003).

[3] A. Georges, *et al.*, Reviews of Modern Physics **68,** 13 (1996).

[4] G. Kotliar and D. Vollhardt, Phys. Today **57**(3), 53 (2004).

[5] A. Sekiyama *et al.*, Phys. Rev. Lett. **93,** 156402 (2004).

[6] K. Kobayashi et al., Appl. Phys. Lett. **83**, 1005 (2003).

[7] Y. Takata et al., Appl. Phys. Lett. **84**, 4310 (2004)

[8] K. Horiba *et al.*, Phys. Rev. Lett. **93,** 236401 (2004).

[9] G. Panaccione *et al.*, Phys. Rev. Lett. **97**, 116401 (2006).

[10] B. L. Chamberland, CRC Crit. Rev. Solid State Mater. Sci. **7,** 1 (1977).

[11] P. Dowben, J. Phys.: Cond. Matter **19**, 310301 (2007).

[12] K. P. Kämper *et al.*, Phys. Rev. Lett. **59,** 2788 (1987).

[13] L. Ranno, A. Barry and J. M. D. Coey, J. Appl. Phys. **81,** 5774 (1997).

[14] F. Hong and J. G. Che, Phys. Rev. Lett. **96**, 167206 (2006).

[15] C. A. Ventrice Jr. et al., J. Phys.: Cond. Matt. **19**, 315207 (2007).

[16] L. Craco, M. S. Laad and E. Müller-Hartmann, Phys.Rev. Lett. **90**, 237203 (2003).

[17] M. A. Korotin *et al.*, Phys. Rev. Lett. **80,** 4305 (1998).

[18] I. I. Mazin, D. J. Singh and C. Ambrosch-Draxl, Phys. Rev. B **59**, 411 (1999).

[19] J. Kunes *et al.*, Phys. Rev. B **65**, 165105 (2002).

[20] A. Toropova *et al.*, Phys Rev. B **71,** 172403 (2005).

[21] A. Yamasaki *et al.*, Phys. Rev. Lett. **98,** 156402 (2007).

[22] S. Suga, Appl. Phys. A **92**, 479 (2008).

[23] A. Biehler *et al.*, Phys. Rev. B **75**, 184427 (2007).

[24] T. Leo *et al.*, Appl. Phys. Lett. **91**, 252506 (2007).

[25] K. Schwarz, J. Phys. F **16**, L211 (1986).

[26] R. J. Soulen *et al.*, Science **282**, 85 (1998).

[27] Y. Ji *et al.*, Phys. Rev. Lett. **86**, 5585 (2001).

[28] Yu. S. Dedkov *et al.*, Appl. Phys. Lett. **80,** 4181 (2002).

[29] M. I. Katsnelson *et al.*, Rev. Mod. Phys. **80**, 315 (2008).

[30] See Supplemental Material for detailed descriptions.

[31] S. Tanuma, C. J. Powell and D. R. Penn, Surf. Interface Anal. **25**, 25 (1997).

[32] M. Budke and M. Donath, Appl. Phys. Lett. **92**, 231918 (2008).

[33] C. F. Chang *et al.*, Phys. Rev. B **71**, 052407 (2005).

[34] M. Taguchi *et al.*, Phys. Rev. B **71**, 155102 (2005).

[35] K. Okada and A. Kotani, J. Phys. Soc. Jpn. **75**, 123703 (2006).

[36] S. Suga *et al.*, New J. Phys. **11**, 073025 (2009).

[37] Y. Takata *et al.*, Phys. Rev. Lett. **101,** 137601 (2008).

[38] Z. V. Pchelkina *et al.*, Phys. Rev. B **75,** 035122 (2007).

[39] M. Donath, Surf. Sci. Rep. **20**, 251 (1994).

[40] M. Budke *et al.*, Rev. Sci. Instrum. **78**, 083903 (2007).

[41] T. Tsujioka *et al.*, Phys. Rev. B **56,** R15509 (1997).

**Figure Captions**

Figure 1: (color online). Valence band PES of epitaxial $CrO_2$(100) films at 20 K for different photon energies in normal emission. The surfaces are as-grown, including a nominal $Cr_2O_3$ surface layer.

Figure 2: (color online). Valence band PES of epitaxial $CrO_2$(100) films for $h\nu = 7942$ eV in normal emission at 150K and 20 K. The red and blue lines are guides to the eye. Near $E_F$ the $CrO_2$ spectrum at 20 K (expanded ×5) is compared to that of a Au film.

Figure 3: (color online). PES of epitaxial $CrO_2$(100) films at 20 K for different photon energies. (a) O 1$s$ core level PES; expanded ×10 for $h\nu = 8180$ eV. (b) Cr 2$p$ core level PES; the Cr 2$p_{3/2}$ state near 576 eV BE exhibits a well-screened feature near 575 eV.

Figure 4: (color online). IPES spectra at room temperature on as-grown and sputter-cleaned (using $Ne^+$ ions [30]) epitaxial $CrO_2$(100) film.

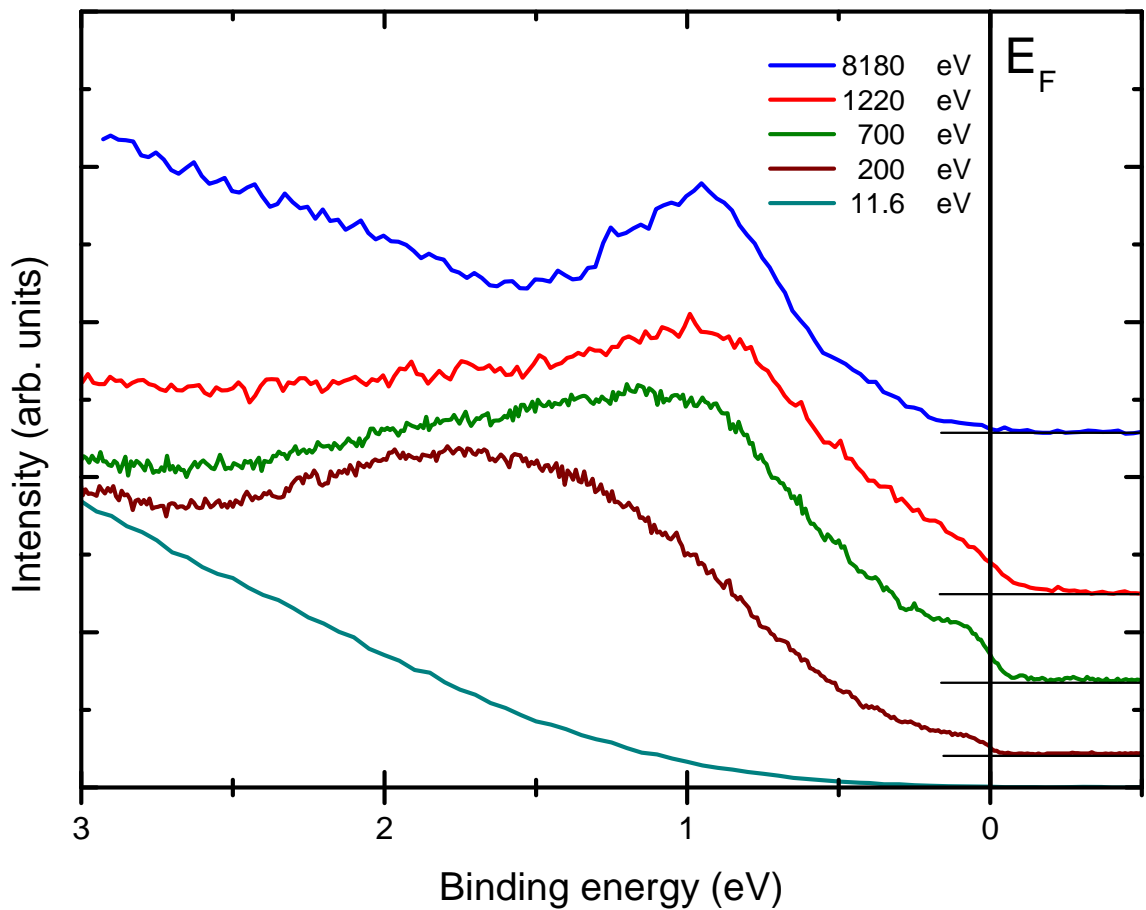

Figure 1.  M. Sperlich *et al.*

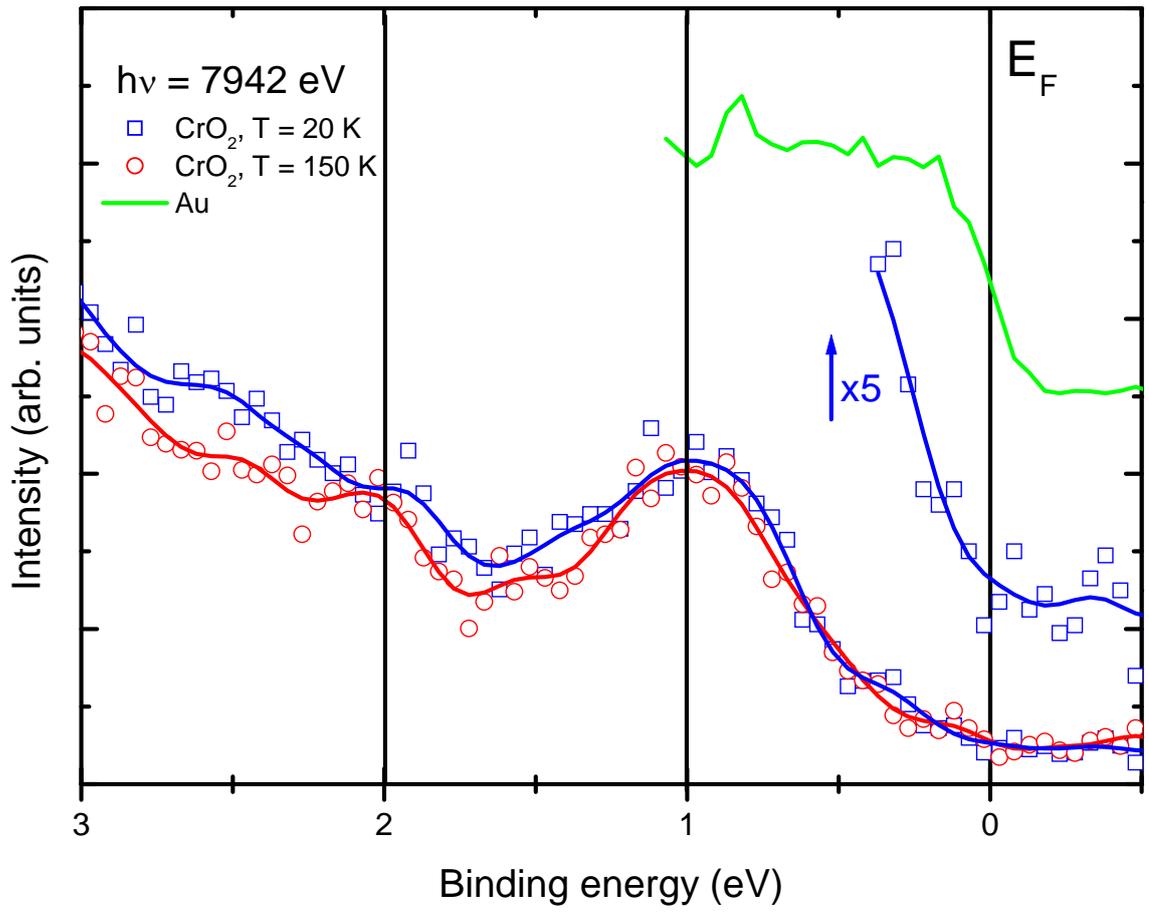

Figure 2. M. Sperlich *et al.*

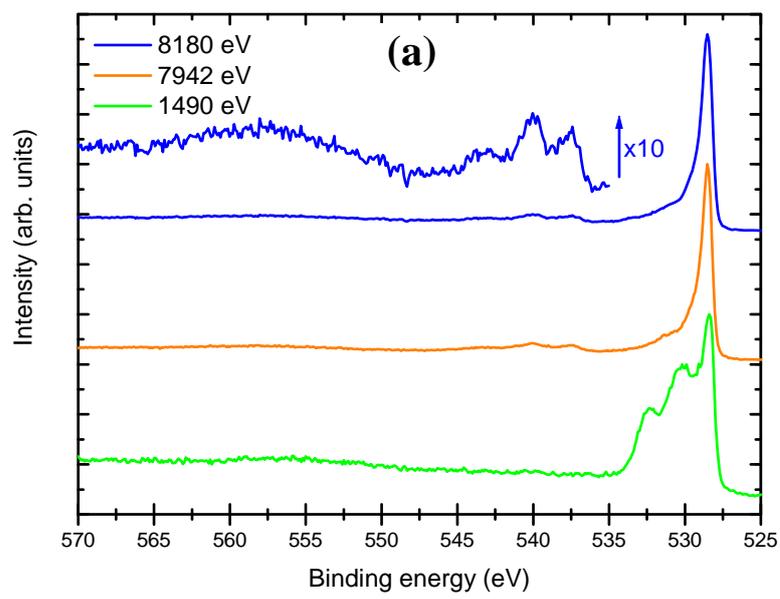

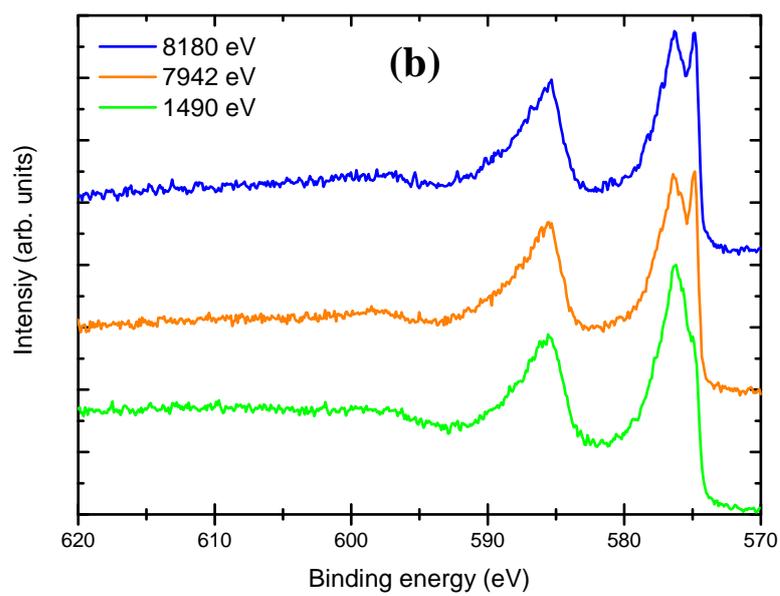

**Figure 3.**   M. Sperlich *et al.*

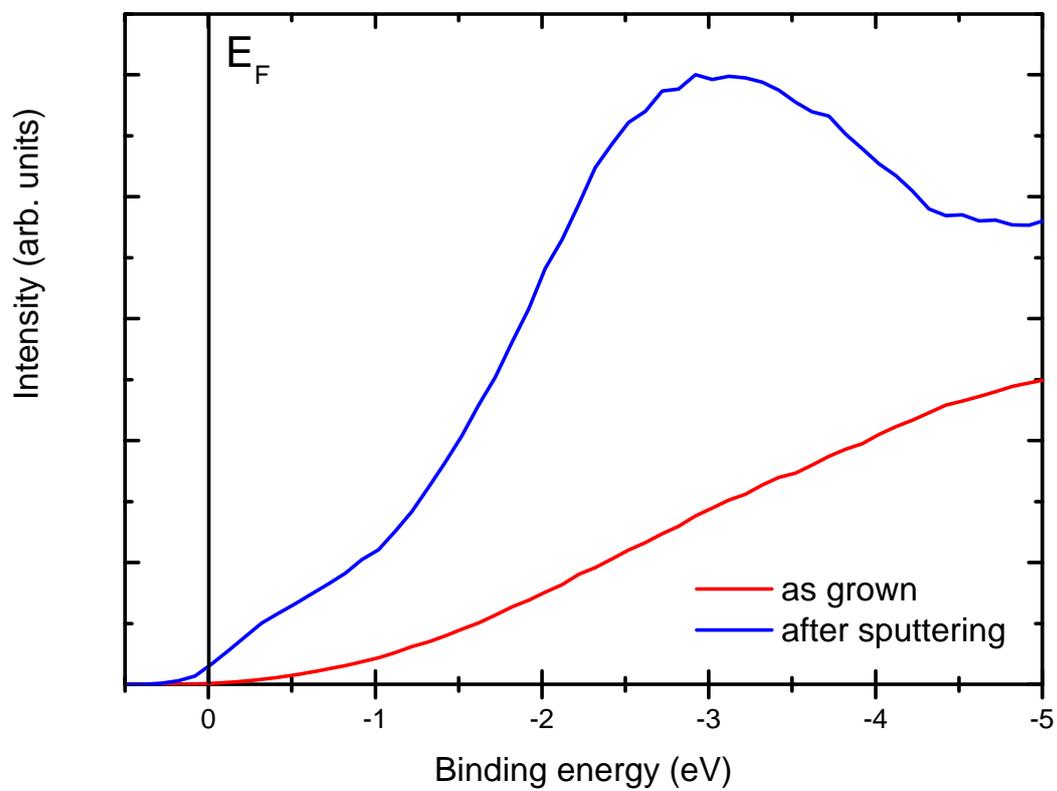

**Figure 4.** M. Sperlich *et al.*

Supplemental Material (SM) for "Intrinsic Correlated Electronic Structure of $CrO_2$ Revealed by Hard X-ray Photoemission Spectroscopy"

M. Sperlich[1], C. König[1], G. Güntherodt[1], A. Sekiyama[2], G. Funabashi[2], M. Tsunekawa[2], S. Imada[2], A. Shigemoto[2], K. Okada[3], A. Higashiya[4], M. Yabashi[4], K. Tamasaku[4], T. Ishikawa[4], V. Renken[5], T. Allmers[5], M. Donath[5] and S. Suga[2]

[1] *II. Physikalisches Institut, RWTH Aachen University, 52074 Aachen, Germany*

[2] *Graduate School of Engineering Science, Osaka University, Toyonaka, Osaka 560-8531, Japan*

[3] *Graduate School of Natural Science and Technology, Okayama University, Okayama 700-8530, Japan*

[4] *SPring-8/Riken, 1-1-1 Kouto, Mikazuki, Sayo, Hyogo 679-8148, Japan*

[5] *Physikalisches Institut, Westfälische Wilhelms-Universität Münster, 48149 Münster, Germany*

## I. METHODS

### A. HAXPES measurements

The HAXPES experiments were performed at the BL19LXU beam line of SPring-8 using a 27 m long in-vacuum planar undulator with 781 periods. A newly developed MBS A1-HE hemispherical electron analyzer (using up to 14 keV) was employed. The long term stability of the high voltage power supply was ± 8 mV. The resolution of the system was checked by the Fermi edge of Au kept at 20 K and found to be better than 55 meV for the pass energy of 50 eV and the slit width of 0.2 mm at $h\nu$ = 7.94 keV. For the present measurement, the resolution was set to 250 meV (FWHM) to overcome the very low photoionization cross section (PICS) of the Cr $3d$ valence states in this material. To optimize the latter, the photoelectron emission direction was chosen parallel to the photon electric-field vector. The measurements were carried out with decreasing the temperature.

B. IPES measurements

The inverse photoemission spectroscopy (IPES) measurements [39] were performed in an ultrahigh vaccum system with a base pressure of less than $3 \cdot 10^{-11}$ mbar. Low-energy electrons (8 to 15 eV) from a GaAs photocathode impinge on the sample. Radiative transitions into lower lying unoccupied electronic states within the sample are monitored via emitted photons. The photons are detected at fixed energy in Geiger-Müller counters acting as band-pass detectors. The mean detection energy and the width of the energy band pass are determined by the combination of the counting gas (photothreshold) and entrance window (transmission cut-off). In our case, we used acetone as counting gas and $CaF_2$ as entrance window at room temperature. This choice results in a detection energy of 9.9 ± 0.15 eV. The overall energy resolution including the energy spread of the electron beam amounts to 350 meV [40]. Because of the high surface sensitivity of IPES the surface of $CrO_2$ has been sputtered by using $Ne^+$ ions with 600 eV kinetic energy at grazing incidence for 66 hours applying 2.4 µA over 2 $cm^2$ sample area.

C. UPES measurements

For the UPES ($h\nu$ = 11.6 eV) experiments in a laboratory system an as-grown surface (Fig. 1) and a sputter-cleaned surface (Supplemental Figure 1) has been used. The latter exhibits a finite emission intensity and a Fermi-type edge at $E_F$, demonstrating the high surface sensitivity of UPES. For sputtering we used the same conditions as for IPES (see above).

II. SAMPLE PREPARATION

The samples were grown by chemical vapor deposition in an oxygen atmosphere on (100)-oriented $TiO_2$ substrates. In a two-zone furnace the temperature for the $CrO_3$ precursor was set to 270°C while the substrates were heated up to 390°C. An oxygen gas flow of about 10 ml/min (increased to 60 ml/min during cooling of the sample) in combination with the exact temperatures provided the best growth parameters in terms of crystal quality and surface roughness. Phase purity and pronounced film texture have been investigated by $\theta$-$2\theta$ x-ray diffraction showing the (200) and (400) reflections of $CrO_2$ and $TiO_2$ and no other $CrO_x$ phases. Low-energy electron diffraction measurements [SM1] showed lattice parameters of (4.3±0.2) Å and (2.7±0.2) Å for the [010] and [001] directions, respectively, which are in good agreement with the tetragonal rutile structure of $CrO_2$ with $a = b = 4.42$ Å and $c = 2.92$ Å [SM2]. Atomic force microscopy measurements showed an rms roughness of about 0.6 nm over an area of $2 \times 2$ $\mu m^2$.

### III. SPECTROSCOPIC DATA

#### A. Valence band spectrum overview (Supplemental Figure 2.)

In the Supplemental Fig. 2 we show the whole valence band spectrum of $CrO_2$ ($E_F \leq BE \leq 14$ eV) for $h\nu = 7942$ eV at 20 K and 150 K. The spectrum shows a main maximum near 6.5 eV BE with a pronounced shoulder near 5 eV BE due to the O 2p derived valence band [17], a small shoulder near 2 eV BE and a sharp peak near 1 eV BE. For reference the Fermi edge of Au is also shown.

#### B. Satellites of O 1s core level (Supplemental Figure 3.)

In order to explain the different satellite features of the O 1s core level, cluster calculations have been carried out based on a simplified two-dimensional multi-band Hubbard model [SM3]. In this model the Cr $3d(3z^2-r^2)$ - O $2p_z$ hybridized states are influenced by the exchange field of ferromagnetically ordered and localized Cr $3d(t_{2g})$ electrons. The atomic configuration in our cluster model is shown in the inset of the Supplemental Figure 3. The Hamiltonian is given by

$$H = \sum_{i,\sigma}(\varepsilon_d - J_{ex}\delta_{\sigma\uparrow})d^+_{i\sigma}d_{i\sigma} + \sum_{j,\sigma}\varepsilon_p p^+_{j\sigma}p_{j\sigma} + \sum_{\langle i,j \rangle,\sigma}V_{pd}(d^+_{i\sigma}p_{j\sigma} + h.c.) + \sum_{\langle j,j' \rangle,\sigma}V_{pp}(p^+_{j\sigma}p_{j'\sigma} + h.c.)$$
$$+ \sum_i U_{dd}d^+_{i\uparrow}d_{i\uparrow}d^+_{i\downarrow}d_{i\downarrow} + \sum_j U_{pp}p^+_{j\uparrow}p_{j\uparrow}p^+_{j\downarrow}p_{j\downarrow} - Q(1-n_{core})\sum_\sigma d^+_{i0}d_{i0}$$

where the first term on the right-hand side belongs to the spin-dependent Cr $3d(3z^2-r^2)$ states and $J_{ex}$ represents the exchange coupling energy between localized Cr $3d(t_{2g})$ and delocalized Cr $3d(3z^2-r^2)$ electrons. $\delta_{\sigma\uparrow}$ denotes Kronecker's symbol, which equals 1 for spin up and 0 for spin down. The Cr $3d(x^2-y^2)$ orbitals are not taken into account explicitly, whereas it is assumed that the effect of Cr $3d(x^2-y^2)$ orbitals is renormalized into the Cr $3d(3z^2-r^2)$ degree of freedom. The second term is due the O $2p_z$ states which couple to Cr $3d(3z^2-r^2)$. The third and fourth terms are nearest-neighbor p-d and p-p hybridizations, respectively. The fifth and sixth terms represent on-site Coulomb repulsion at Cr and O sites, respectively. The last term is the core-hole attractive potential, where $n_{core}$ is the number operator for the O 1s (spin-less) core electron at site 0. The parameter values used in our model calculation are as follows: $\varepsilon_d - \varepsilon_p = 3.5$ eV, $pd\sigma = -2$ eV, $pp\sigma = 1$ eV, $pp\pi = -0.3$ eV, $U_{dd} = 6$ eV, $U_{pp} = 3$ eV, $Q = 6$ eV, $J_{ex}$ = 2 eV. These values were taken from $TiO_2$, $VO_2$ [SM4] and cuprates [36, SM5]. From the latter the value of $Q$ = 6 eV is taken. The O $1s$ spectral function is calculated under the sudden approximation by using the Lanczos method. The obtained line spectrum is convoluted by using a Lorentz function of width 0.3 eV (HWHM). The result shown in the Supplemental Figure 3 reproduces well the Cr $3d(e_g)$ - O $2p$ charge transfer (CT) satellites at 9 eV and 11 eV. Their splitting is due to exchange interaction between Cr $3d$ states. For the description of the third peak near 13.5 eV, which cannot be reproduced within the present theory, one has to take into account the details of the conduction bands. Besides these satellites, there are weak structures predicted near 3 - 6 eV, which are mainly caused by O $2p$ - O $2p$ CT. Our conclusions for the satellites near 9eV and 11 eV are consistent with satellite structure found in the O $1s$ X-ray photoemission spectroscopy (XPS) of cuprates, where this is caused by a CT from the O $2p$ orbitals to the Cu $3d(x^2-y^2)$ states [35]. The line shape is found to reflect the DOS of the unoccupied electronic states just above $E_F$.

### C. Absence of well screened peak of Cr $2p_{1/2}$ level

The well screened feature found for the Cr $2p_{3/2}$ level is not resolved for the Cr $2p_{1/2}$ level presumably due to a larger manifold of configurational states compared to the 3/2-state. This may result in a larger lifetime broadening of the 1/2-state. In 3d transition metal elements the width of the $2p_{1/2}$ level may be larger than that of the $2p_{3/2}$ level due to L2-L3-M45 Coster-Kronig transitions, which has been reported for the Cu 2p line shape analysis of an organic conductor [SM6]. A well-screened peak is also absent in the Cr $2p_{1/2}$ spectra of the $Cr^{4+}$ based $CaCrO_3$ for $hv$ = 1.2 keV [SM7] and in the Mn $2p_{1/2}$ spectra for $hv$ = 5.95 keV in the metallic phase of $La_{1-x}Sr_xMnO_3$ [8]. Both cases have been modeled by correlated electron structures of the charge-transfer type.

## IV. RECOIL SHIFTS
### A. $p$- vs $d$-states weighted average of recoil shifts

The recoil shift of O($2p$) vs Cr($3d$) hybridized states is estimated by the (6 vs 10) degeneracy-weighted average of the estimated recoil shifts $E_R^{H-S}$ of the O($1s$) and Cr($2p$) core levels, yielding a value of 140 meV. Analogously, a spin-polarized $2p\uparrow$ vs $3d\uparrow$ degeneracy (3 vs 5) weighted average at $E_F$, considering the partial weight of the density of states of 0.23 for $2p\uparrow$ vs 0.77 for $3d\uparrow$ states [17], yields a value of 110 meV.

### B. Experimental O $1s$ and Cr $2p_{3/2}$ recoil shifts (Supplemental Figure 4.)

In the Supplemental Figure 4 the core level spectra of the O 1s and Cr $2p_{3/2}$ states are shown each for HAXPES ($h\nu$ = 7942 eV) and SXPES ($h\nu$ = 1490 eV), the differences in BE of which amount to 180 meV for O 1s and 70 meV for Cr $2p_{3/2}$.

   C.  Resolution of recoil shifts (Supplemental Figure 5.)

A test of resolving recoil shifts of 100 meV in HAXPES with 250 meV resolution is shown in the Supplemental Figure 5 for two Gaussian lineshapes which have been broadened by 250 meV. The two peaks can obviously be discerned.

   V.  ORBITAL DEPENDENT PHOTOIONIZATION CROSS SECTIONS (PICS) in SXPES and HAXPES

The strongly suppressed intensity in the *p-d* hybridized valence band spectrum for hν = 8180 eV and $E_F \leq$ BE $\leq$ 0.2 eV compared to SXPES (Fig. 1) is primarily not due to a stronger dropping of the calculated PICS of the atomic subshell O *2p* compared to the Cr 3d subshell. Both subshells track each other for hν > 500 eV with minor differences [SM8-SM10]. At 1500 eV the cross section of O *2p* has decreased by roughly a factor of 4 compared to that of Cr *3d* [SM8], i.e. the O *2p* states do not disappear much stronger than the Cr *3d* states with increasing X-ray photon energy. Moreover, from 1 to 8 keV the relative Cr3d/O2p PICS hardly changes from 2.9 to 2.6 [SM9]. A slightly stronger decrease from 3.1 (for 1 keV) to 1.9 (for 8 keV) is found in Ref. [SM10]. On the contrary, the d-derived bulk (coherent) quasiparticle peak near 1 eV BE is enhanced with hν approaching 8 keV according to the increased bulk sensitivity. The 1.75-eV peak for hν = 200 eV in Fig. 1 is mostly due to the surface-sensitive d state [28] which is more localized and rather close to the LHB in the bulk.

Supplemental Material (SM) References

[SM1] Yu. S. Dedkov *et al.* Phys. Rev. B **72**, 060401 (2005).

[SM2] B. J. Thamer, R. M. Douglass and E. Staritzky, J. Am. Chem. Soc. **79**, 547 (1957).

[SM3] K. Okada *et al.*, unpublished.

[SM4] T. Uozumi, K. Okada and A. Kotani, J. Phys. Soc. Jpn. **62**, 2595 (1993).

[SM5] A. K. McMahan, R. M. Martin and S. Satpathy, Phys. Rev. B **38**, 6650 (1988).

[SM6] I. H. Inoue *et al.*, Phys. Rev. B **45**, 5828 (1992).

[SM7] P. A. Bhobe *et al.*, Phys. Rev. B **83**, 165132 (2011).

[SM8]  J. J. Yeh and I. Lindau, At.Data Nucl. Data Tables **32**, 1 (1985).

[SM9]  J. H. Scofield, Lawrence Livermore Lab. Rep. UCRL-51326 (1973).

[SM10]  M. B. Trzhaskovskaya *et al.*, At. Data Nucl. Data Tables **82**, 257 (2002); ibid **92**, 245 (2006).

Supplemental Figure Captions

Supplemental Figure 1: (color online). UPES ($h\nu$ = 11.6 eV) spectrum of a sputtered $CrO_2$(100) film at 300 K; sputtering conditions as for IPES in Fig. 4 (see also above).

Supplemental Figure 2: (color online). Valence band photoemission spectra of epitaxial $CrO_2$(100) films at 150K and 20 K for hν = 7942 eV and $E_F$ ≤ BE ≤ 14 eV in normal emission. No significant temperature dependence is found, except for intensity for 4 eV ≤ BE ≤ 7 eV. The photoemission intensity near $E_F$ is compared to that of a Au film.

Supplemental Figure 3: (color online). Cluster calculation of the oxygen O 1$s$ level and its satellite structure (see text). The inset shows the atomic configuration used in the model. The crystallographic c axis is perpendicular to the x'-z plane. Blue and open circles denote Cr and O atoms, respectively.

Supplemental Figure 4: (color online). O 1$s$ and Cr $2p_{3/2}$ core level spectra for HAXPES ($h\nu$ = 7942 eV) and SXPES ($h\nu$ = 1490 eV). In the SXPES line shape of the O 1$s$ core level some surface contributions on the higher BE side are estimated by the spectral line shape between the peak and the lower BE tail as indicated by the dashed black line. The solid black line in the HAXPES Cr $2p_{3/2}$ level indicates its experimentally estimated shift with respect to its corresponding SXPES position. This estimate is difficult because of the 2p spectral weight consisting of multiple structures due to nonlocal screening and atomic multiplets whose weight changes considerably between bulk and surface. We consider a value of 70 meV for the recoil shift of the Cr $2p_{3/2}$ level as a reasonable approximation (see text).

Supplemental Figure 5: (color online). Resolution test of recoil shifts of 100 meV for two Gaussian lineshapes broadened to 250 meV FWHM.

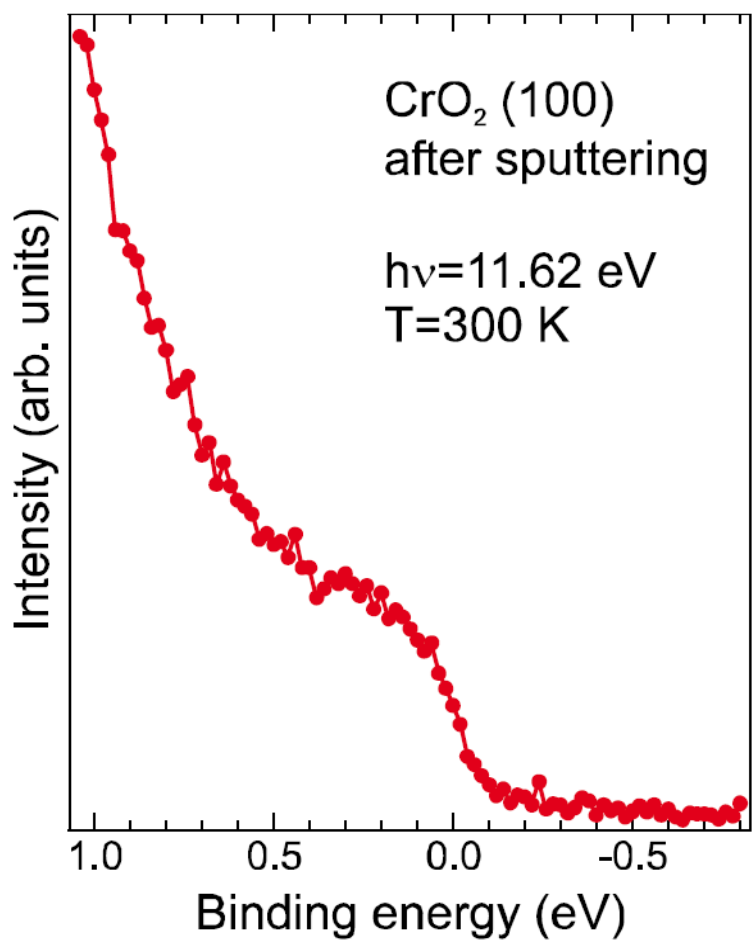

**Supplemental Figure 1.   M. Sperlich *et al.***

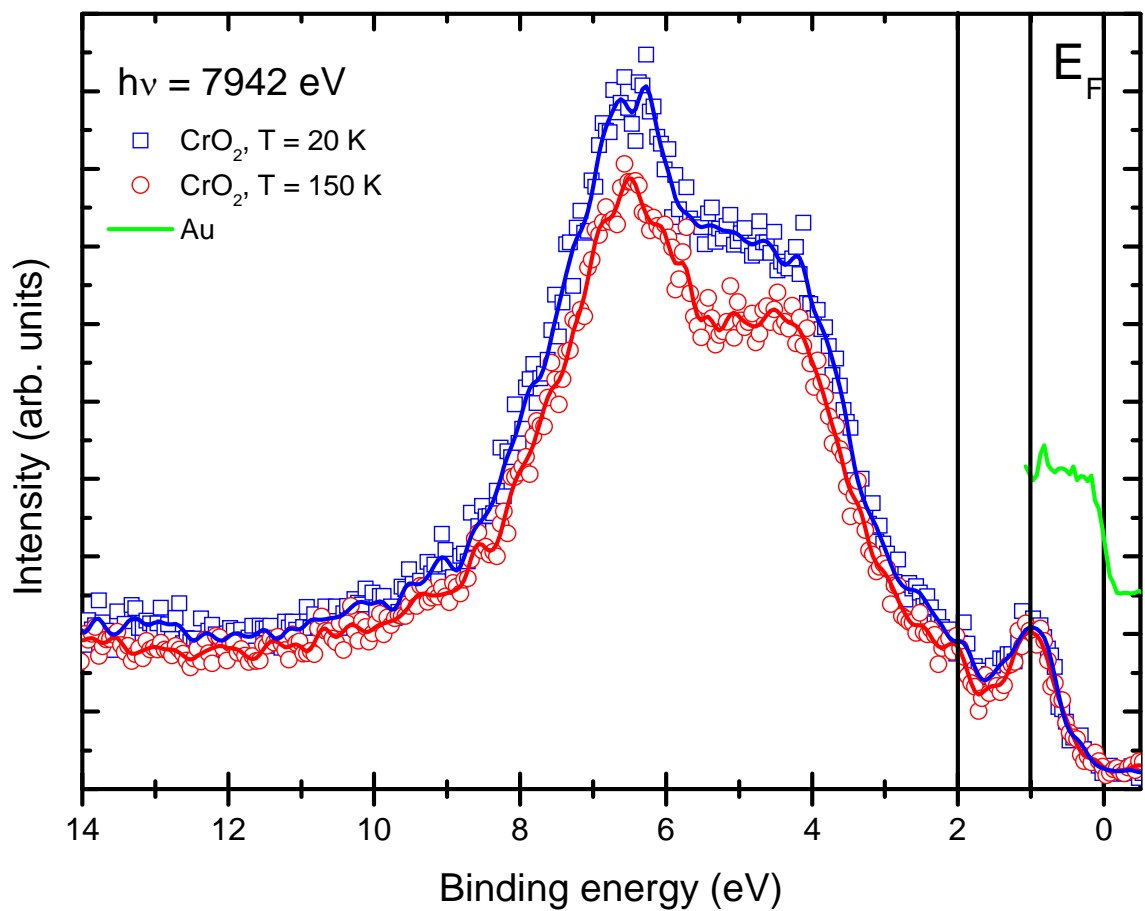

**Supplemental Figure 2.** M. Sperlich *et al.*

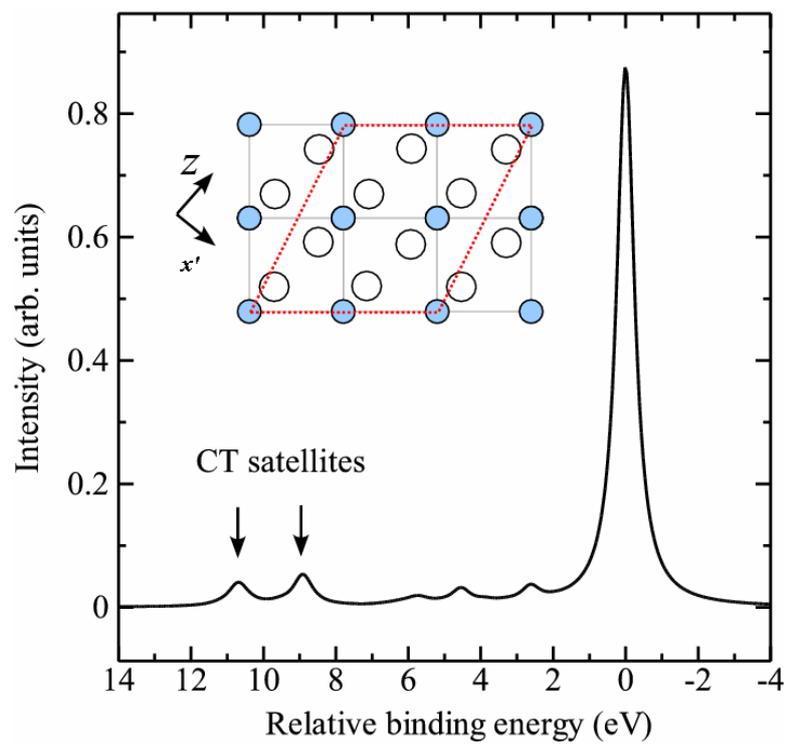

**Supplemental Figure 3.   M. Sperlich *et al.*

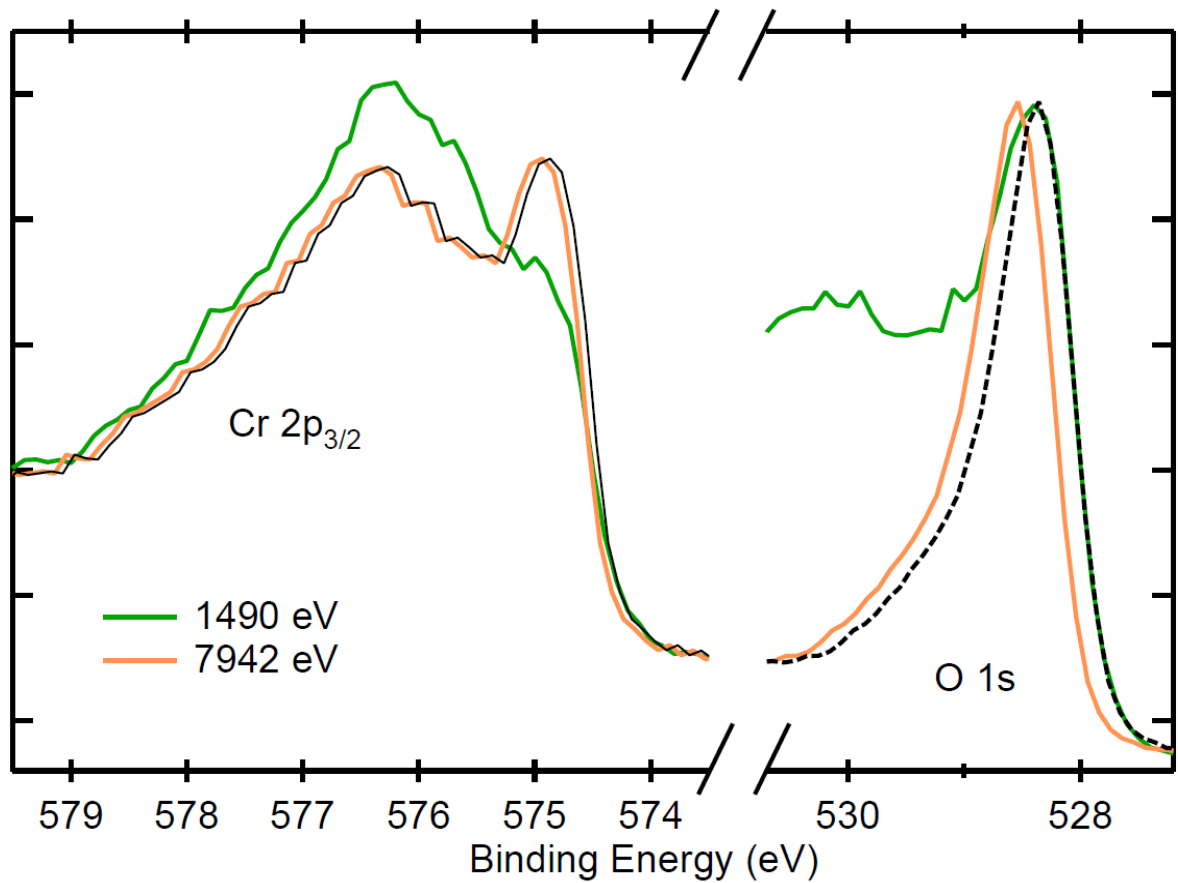

Supplemental Figure 4.  M. Sperlich *et al.*

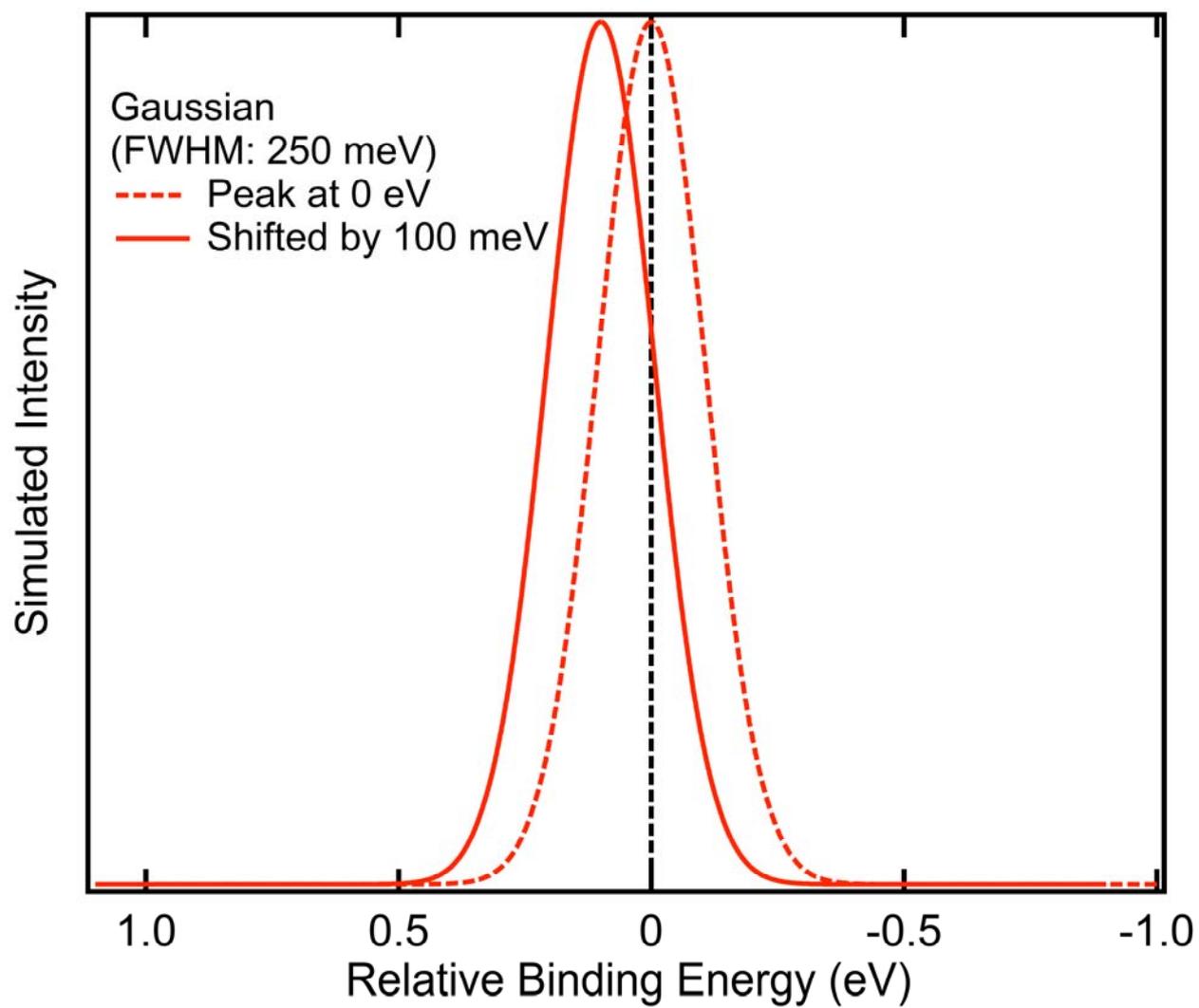

**Supplemental Figure 5.   M. Sperlich** *et al.*